\documentclass[10pt, twocolumn]{IEEEtran}

\usepackage{xspace,amsmath,amssymb,epsfig,syntonly,cite}
\usepackage{verbatim}

\newtheorem{lemma}{\hspace{-11pt}\bf Lemma}
\newtheorem{algorithm}{\hspace{-11pt}\bf Algorithm}

\newtheorem{proposition}{\hspace{-11pt}\bf Proposition}

\newtheorem{remark}{\hspace{-11pt}\bf Remark}

\usepackage{epstopdf}
\usepackage{color}
\long\def\symbolfootnote[#1]#2{\begingroup
\def\thefootnote{\fnsymbol{footnote}}
\footnote[#1]{#2}\endgroup}

\psfull

\begin{document}
\title{Joint Base Station Activation and Coordinated Downlink Beamforming for HetNets: Efficient Optimal and Suboptimal Algorithms}
\author{
   {Siguo Bi},
   {Zhaoxi Fang},
   {Xiaojun~Yuan},~\IEEEmembership{Senior Member,~IEEE},
   {and Xin~Wang},~\IEEEmembership{Senior Member,~IEEE}      \\

    \thanks{The work in this paper was supported by the National Natural Science Foundation of China Grants No. 61671154, the National Key Research and Development Program of China Grant 2017YFB0403402, and the Innovation Program of Shanghai Municipal Science and Technology Commission Grant 17510710400. Part of the work was presented at Globecom 2015 [1].

S. Bi and X. Wang are with the Shanghai Institute for Advanced Communication and Data Science, Key Laboratory for Information Science of Electromagnetic Waves (MoE), Department of Communication Science and Engineering, Fudan University, Shanghai, China (email: xwang11@fudan.edu.cn).

    Z. Fang is with the Faculty of Electronic and Information Engineering,
Zhejiang Wanli University, Ningbo, China, e-mail:
zhaoxifang{\rm\char64}gmail.com.

 X. Yuan is with  the Center for Intelligent Networking
and Communications (CINC), the University of Electronic Science and Technology of China, Chengdu, China, email: xjyuan@uestc.edu.cn.
}
%
%
%
%
}

\markboth{} {}

\maketitle

\setcounter{page}{1}
\begin{abstract}
In cellular heterogeneous networks (HetNets),  a number of distributed base stations cooperatively provide services to multiple mobile users. This paper addresses joint base-station activation and coordinated beamforming design for downlink transmission in HetNets. To this end, a mixed integer program is formulated to optimize the total power consumption of a HetNet. A novel approach based on Benders' decomposition is then put forth to obtain the optimal solution for the problem of interest with guaranteed convergence. Building on our new formulation, a dual-subgradient algorithm is also proposed to find an approximate solution  in polynomial time. The proposed approaches can be generalized to more general setups, including robust beamforming designs under channel uncertainty, and coordinated beamforming for multi-cell scenarios.

\end{abstract}


\section{Introduction}
To meet the explosively growing demand for mobile date services, the current cellular wireless networks are evolving into heterogeneous
networks (HetNets) consisting of many small cells \cite{Hwa13,Fan16,Fan17}. It has been shown that the HetNets with densely deployed base stations can have great advantages over the traditional cellular architecture comprising a few high-power base stations (BSs)\cite{Dam11}.

In HetNets, the coexistence of many close BS transmitters can introduce severe mutual interference. To overcome this issue, coordinated transmissions based architectures, such as the coordinated multi-point process (CoMP), have been proposed for next-generation cellular networks \cite{Tau17, Ali17, Ali18, Par15, Bas16}. To fully exploit their potentials, coordinated beamforming and BS cooperation were investigated in \cite{Tol11, Dal10, Zha09, Liu11, Ng10, Che13, Hong13}. The growing number of small cells has also invoked the interest of investigating the energy efficiency of HetNets. The related works on energy efficiency of cellular networks have been investigated in \cite{Ge181, Ge182, Xiang13}. Due to dense deployment of the small-cell BSs, the electricity cost has become a substantial part of the operational expenditure for cellular service providers. In addition, $\text{CO}_2$ emissions by cellular networks has contributed a significant portion of the global ``carbon footprint" \cite{Oh11}. Driven by these economic and ecological concerns, energy-saving coordinated beamforming schemes have been developed in \cite{Dal10, Che13, Shi13, Cao10, Rao13}.

The spectral- and energy-efficiency can be substantially improved by the coordinated transmissions with full BS cooperation, at the cost of substantially increased operational  and backhaul communication overheads. To balance the benefits and coordination overheads, a mixed-integer conic programming problem was formulated to pursue joint BS activation/clustering and coordinated beamforming schemes that minimize the total power consumption for CoMP downlink \cite{Che13}. A branch-and-cut scheme was developed to approach globally optimal solution  with a very high complexity, while heuristic inflation and deflation procedures were put forth to find an approximate solution in polynomial time, at the cost of substantial performance loss. Based on a sparsity pursuit paradigm, other sub-optimal algorithms were also developed to address the spectral- and/or energy-efficiency for HetNets\cite{Hong13, Liao14, Shi13, Kua14}. A common theme of these approaches is to add a proper sparsity regularizer into the objective functions to render the desired group sparsity structure of the resultant coordinated beamforming vectors across the BSs.

In this paper, we develop novel approaches to pursue efficient joint BS activation and coordinated downlink beamforming design with affordable complexity. To this end, a new mixed-integer programming problem is formulated, where the group sparsity constraints are imposed in an explicit and quantitative manner, rather than implicitly through the addition of sparsity regularizers as in \cite{Hong13, Liao14, Shi13, Kua14}.
The judiciously formulated mixed integer program has a separable structure in the binary variables of BS activation indices and the continuous variables of beamforming vectors. Relying on the generalized Benders' decomposition approach \cite{Geo72}, a master problem and an associated coordinated beamforming design subproblem are formulated from the decomposition of the original problem. By solving a series of relaxed master programs and the associated convex subproblems, the proposed Benders' decomposition based approach can find the global optimum with only a finite number of iterative computation. Note that a recent paper has also used the Benders' decomposition approach to address joint BS association and power control for HetNets \cite{Qian13}. However, \cite{Qian13} only relied on the original Benders' decomposition to consider single-antenna BSs with limited coordination (via power control, without coordinated beamforming among BSs). Capitalizing on the generalized Benders' decomposition, our approach is able to address more sophisticated coordinated beamforming design in multi-antenna BS scenario. In addition, based on our novel formulation, we also develop a low-complexity dual-subgradient based method to find an approximate (near-optimal) solution for our problem of interest in polynomial time.

Our contributions are summarized as follows.
\begin{enumerate}
\item We explicitly formulate a new mixed-integer programming problem with group sparsity constraints in an explicit form. The formulation leads to a naturally separable structure in the binary variables and the continuous variables, which is well suited for implementation of Benders' decomposition.
\item We propose a novel generalized Benders' decomposition method to obtain the globally optimal solution with  affordable complexity.
\item We also develop a low-complexity dual-subgradient based method to find a near-optimal solution in polynomial time.
\item We further generalize our proposed framework to robust beamforming designs accounting for CSI errors, and to multi-cell HetNet setups performing partial coordinated
transmissions.
\end{enumerate}

We organize the remainder of this paper as follows. Section II describes the system models. The proposed Benders' decomposition approach to joint BS activation and beamforming design for coordinated transmission is proposed in Section III. A dual-subgradient based method is further developed in Section IV. Extensions of the proposed approaches to more general setups are outlined in Section V. Section VI provides simulation results to corroborate the superior performance
of the proposed schemes over the existing alternatives.

{\em Notations}: Boldface fonts denote vectors or matrices, calligraphy fonts denote sets, $\mathbb{C}^{L\times K}$ and $\mathbb{R}^{L\times K}$ denote the $L$-by-$K$ dimensional complex and real space; $(\cdot)^T$ denotes transpose, and $(\cdot)^H$ denotes conjugate transpose; $\text{diag}(P_1, \ldots, P_L)$ denotes a diagonal matrix with $P_1, \ldots, P_L$ as the diagonal entries; $\text{Re}(\cdot)$ and $\text{Im}(\cdot)$ denote the real and imaginary parts of a complex scalar; $| \cdot |$ denotes norm of a complex scalar, and $\| \cdot \|$ the Euclidean norm of a complex vector; $\boldsymbol{0}$ denotes all-zero vectors; $\boldsymbol{I}$ denotes the identity matrix; the vector inequalities are defined element-wise; $\text{tr}(\boldsymbol{X})$ and $\text{rank}(\boldsymbol{X})$ denote the trace and rank operators for matrix $\boldsymbol{X}$, respectively; $\boldsymbol{X} \succeq 0$ means that a square matrix $\boldsymbol{X}$ is positive semi-definite.

\section{System Modeling}
\begin{figure}
\centering \epsfig{file=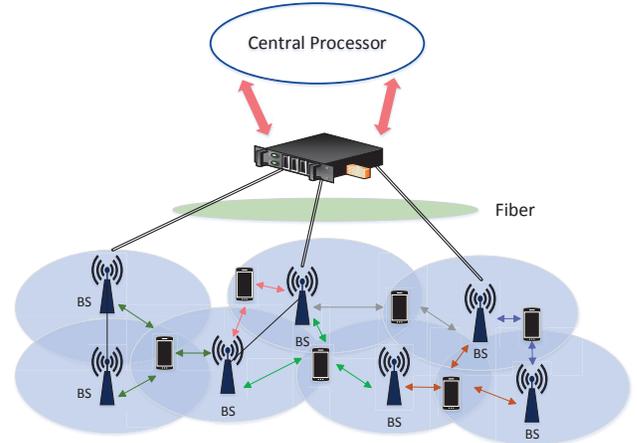,
width=0.50\textwidth}
\caption{A cellular heterogeneous network. }
\label{F:model} 
\end{figure}

Consider a downlink ``single-cell'' HetNet scenario where a set of
distributed BSs ${\cal L}:=\{1,\ldots, L\}$ transmit to a set of
users ${\cal K}:=\{1,\ldots, K\}$ \cite{Hong13, Liao14}; see Fig. 1.
Generalization to multi-cell HetNet will be outlined in Section IV-B. Suppose that each BS is equipped with $N_l \geq 1$ antennas, while each user has only a single antenna. A central entity, which has the knowledge of all the users' data and  global channel station information (CSI), coordinates the cooperative communications among the BS within the cell, through a low-latency backhaul.

As the number of distributed BSs grows large, energy efficiency becomes a key issue. For an actively transmitting BS, a significant portion of power is required for backhaul communications and signal processing, etc. On the other hand,  an inactive BS can be turned into a sleep mode to cut down the aforementioned implementation power consumption. Denote the implementation power with ``active'' mode as $\pi_l$, and that with the ``sleep'' mode as $\rho_l$. Typically  $\pi_l$ is significantly larger than $\rho_l$ in practice. For convenience, we can simply assume $\pi_l>0$ Watts and $\rho_l=0$ Watt without affecting the intended design \cite{Shi13}.


Since $\pi_l$  is non-negligible, we may turn off some active BSs to minimize the total power consumption. Hence, the central controller needs to optimally determine the (sub-)set of active BSs and the beamforming vectors for active BSs subject to various physical constraints.

Denote by $\boldsymbol{h}_{lk} \in \mathbb{C}^{N_l \times 1}$ the channel vector and by $\boldsymbol{w}_{lk}\in \mathbb{C}^{N_l \times 1}$ the transmit-beamforming vector from BS $l$ to user $k$, $\forall l$, $\forall k$. Define ${\cal L}_A \subseteq {\cal L}$ as the subset of active BSs. The signal transmitted from an active BS to user $k$ is then

\begin{equation}
    \boldsymbol{x}_l = \sum_{k=1}^K \boldsymbol{w}_{lk} s_k, \quad \forall l \in {\cal L}_A
\end{equation}
where $s_k$ is the data signal. Assume that $s_k$ is a complex random variable with zero mean and unite variance. The received signal at user $k$ is then
\begin{equation}\label{eq.yk}
    y_k =\sum_{l \in {\cal L}_A} \boldsymbol{h}_{lk}^H \boldsymbol{w}_{lk} s_k + \sum_{i \neq k} \sum_{l \in {\cal L}_A} \boldsymbol{h}_{lk}^H \boldsymbol{w}_{ik} s_i + z_k
\end{equation}
where $\sum_{l \in {\cal L}_A} \boldsymbol{h}_{lk}^H \boldsymbol{w}_{lk} s_k$ is the desired signal for user $k$, $\sum_{i \neq k} \sum_{l \in {\cal L}_A} \boldsymbol{h}_{lk}^H \boldsymbol{w}_{ik} s_i$ is the inter-user interference, and $z_k$ denotes the additive complex Gaussian noise with zero mean and variance $\sigma_k^2$.

Based on (1), we can express the signal-to-interference-plus-noise-ratio (SINR) at user $k$ as
\begin{equation}
    \text{SINR}_k= \frac{|\sum_{l \in {\cal L}_A} \boldsymbol{h}_{lk}^H \boldsymbol{w}_{lk}|^2}{\sum_{i\neq k} (|\sum_{l \in {\cal L}_A} \boldsymbol{h}_{lk}^H \boldsymbol{w}_{ik}|^2) + \sigma_k^2}.
\end{equation}

Let $\gamma_k$ denote the SINR target per user, $\forall k$ $\in$ $(1,\ldots, K)$. The power optimization problem can be formulated as follows.
\begin{subequations}\label{eq.p0}
\begin{align}
& \min_{(\{\boldsymbol{w}_{lk}\},\; {\cal L}_A)} \sum_{l \in {\cal L}_A} \; \sum_{k=1}^K \|\boldsymbol{w}_{lk}\|^2 + \sum_{l \in {\cal L}_A} \pi_l \label{eq.p0a} \\
 \text{s.t.}~~& \frac{|\sum_{l \in {\cal L}_A} \boldsymbol{h}_{lk}^H \boldsymbol{w}_{lk}|^2}{\sum_{i\neq k} (|\sum_{l \in {\cal L}_A} \boldsymbol{h}_{lk}^H \boldsymbol{w}_{ik}|^2) + \sigma_k^2} \geq \gamma_k, \;\; \forall k, \label{eq.p0b}\\
  & \sum_{k=1}^K \|\boldsymbol{w}_{lk}\|^2 \leq P_l, \quad \forall l \in {\cal L}_A \label{eq.p0c} 
\end{align}
\end{subequations}
where $P_l$ denotes the maximum transmit-power allowed per BS $l$.

\section{Joint Design of BS Activation and Coordinated Beamforming}

Power minimization problem (\ref{eq.p0}) in Sec.
\uppercase\expandafter{\romannumeral2} is actually a more generalized optimization problem  of the one in \cite{Liao14}.\footnote{If each BS has the same implementation power, i.e., $\pi_l= 1/\theta$, $\forall l$, then the second term in the objective function: $\sum_{l \in {\cal L}_A} \pi_l \equiv 1/\theta \|\{\|\boldsymbol{w}_{l\cdot}\|^2\}_{l \in {\cal L}}\}\|_0$, where $\boldsymbol{w}_{l\cdot}:=[\boldsymbol{w}_{l1}^T, \ldots, \boldsymbol{w}_{lK}^T]^T$. Then (\ref{eq.p0}) becomes the same as the ``single-cell'' version of problem (7) in \cite{Liao14}.} The problem is not convex and generally NP-hard. To obtain an approximate solution, \cite{Shi13, Hong13, Liao14} relied on group sparsity based relaxations to develop heuristic solvers; yet, global optimum is not guaranteed. In this section, we firstly show  how the problem (\ref{eq.p0}) can be reformulated into a mixed-integer program. Building on such a reformulation, we then develop an efficient algorithm to obtain a globally optimal solution based on Benders' decomposition.

\subsection{Mixed Integer Programming Formulation}

Introduce a binary vector $\boldsymbol{a}:=[a_1, \ldots, a_L]^T$ where $a_l \in \{0,1\}$, $\forall l$, indicates whether the BS $l$ is active ($a_l=1$) or not ($a_l=0$). Let  $\boldsymbol{\pi}:=[\pi_1, \ldots, \pi_L]^T$ collect the implementation powers for the BSs. Problem (\ref{eq.p0}) can be rewritten  as
\begin{subequations}\label{eq.prob}
\begin{align}
& \min_{(\{\boldsymbol{w}_{lk}\},\; \boldsymbol{a})} \sum_{l =1}^L \; \sum_{k=1}^K \|\boldsymbol{w}_{lk}\|^2 + \boldsymbol{a}^T\boldsymbol{\pi} \label{eq.proba} \\
 \text{s.t.}~~& \frac{|\sum_{l=1}^L \boldsymbol{h}_{lk}^H \boldsymbol{w}_{lk}|^2}{\sum_{i\neq k} (|\sum_{l=1}^L \boldsymbol{h}_{lk}^H \boldsymbol{w}_{ik}|^2) + \sigma_k^2} \geq \gamma_k, \;\; \forall k \label{eq.probb}\\
  & \sum_{k=1}^K \|\boldsymbol{w}_{lk}\|^2 \leq a_l P_l, \quad \forall l \label{eq.probc}\\
  & a_l \in \{0,1\}, \;\; \forall l.
\end{align}
\end{subequations}
In our judicious formulation (4), the summations in the objective function (\ref{eq.proba}) and the constraints (\ref{eq.probb}) are over all $l \in {\cal L}$. On the other hand, the beamforming weights $\{\boldsymbol{w}_{lk}, \forall k\}$ are forced to zero in case $a_l=0$, i.e., $\forall l \notin {\cal L}_A$ , through the constraint (4c). 

The non-convex constraint (\ref{eq.probb}) can be reformulated into a convex form as follows. Let $\boldsymbol{h}_k:=[\boldsymbol{h}_{1k}^T, \ldots, \boldsymbol{h}_{Lk}^T]^T$ and $\boldsymbol{w}_k:=[\boldsymbol{w}_{1k}^T, \ldots, \boldsymbol{w}_{Lk}^T]^T$. It can be easily seen that SINRs would not be affected by adding an arbitrary phase rotation to the beamforming vectors $\boldsymbol{w}_k$. By choosing a phase such that $\boldsymbol{h}_k^H \boldsymbol{w}_k$ is real and nonnegative, we can rewrite the SINR constraints (\ref{eq.probb}) into a convex second-order cone (SOC) form \cite{Wie06}:
\begin{equation}
    \sqrt{\sum_{i \neq k} |\boldsymbol{h}_k^H \boldsymbol{w}_i|^2 + \sigma_k^2} \leq \frac{1}{\sqrt{\gamma_k}} \text{Re}(\boldsymbol{h}_k^H \boldsymbol{w}_k), \;\;\; \text{Im}(\boldsymbol{h}_k^H \boldsymbol{w}_k) =0.
\end{equation}
Define
\begin{equation}
    \boldsymbol{B}_l:=\text{diag}\Bigl(\underbrace{0, \ldots, 0}_{\sum_{n=1}^{l-1}N_n}, \underbrace{1, \ldots, 1}_{N_l}, \underbrace{0, \ldots, 0}_{\sum_{n=l+1}^{L}N_n}\Bigr).
\end{equation}
With $\boldsymbol{W}:=\{\boldsymbol{w}_k , \forall k\}$, the problem (\ref{eq.prob}) becomes:
\begin{subequations}\label{eq.p1}
\begin{align}
& \min_{(\boldsymbol{W},\; \boldsymbol{a})} \sum_{k=1}^K \boldsymbol{w}_k^H \boldsymbol{w}_k + \boldsymbol{a}^T\boldsymbol{\pi} \label{eq.p1a} \\
 \text{s.t.}~~& \sqrt{\sum_{i \neq k} |\boldsymbol{h}_k^H \boldsymbol{w}_i|^2 + \sigma_k^2} \leq \frac{1}{\sqrt{\gamma_k}} \text{Re}(\boldsymbol{h}_k^H \boldsymbol{w}_k), \nonumber \\
 & \text{Im}(\boldsymbol{h}_k^H \boldsymbol{w}_k) =0, \quad \forall k \label{eq.p1b}\\
  & \sum_{k=1}^K \boldsymbol{w}_k^H \boldsymbol{B}_l \boldsymbol{w}_k \leq a_l P_l, \quad \forall l \label{eq.p1c} \\
  & a_l \in \{0,1\}, \;\; \forall l\label{eq.p1f}
\end{align}
\end{subequations}

If we relax the binary variable $a_l \in \{0,1\}$ to a real variable with $0 \leq a_l \leq 1$, the problem then turns into a computationally solvable convex SOC program (SOCP) \cite{Boyd04}. The branch-and-bound method can rely on solving a series of such SOCPs to compute the global optimum for (\ref{eq.p1}). However, the computational complexity with such a method can be formidably  high as the number of optimization variables grows.


\subsection{Benders' Decomposition}

We next propose a Benders' decomposition method to obtain the globally optimal solution for (\ref{eq.p1}) with affordable complexity. Define functions $f_1(\boldsymbol{W}):=\sum_{k=1}^K \boldsymbol{w}_k^H \boldsymbol{w}_k$, and $f_2(\boldsymbol{a}):=\boldsymbol{a}^T\boldsymbol{\pi}$. Define vector functions $\boldsymbol{G}_1(\boldsymbol{W}):=\{\sum_{k=1}^K \boldsymbol{w}_k^H \boldsymbol{B}_l \boldsymbol{w}_k, \forall l\}$ and $\boldsymbol{G}_2(\boldsymbol{a}):=\{-a_l P_l, \forall l\}$. Further, let
\begin{align}
    f(\boldsymbol{W},\boldsymbol{a})& :=f_1(\boldsymbol{W})+f_2(\boldsymbol{a}), \label{eq.f} \\
    \boldsymbol{G}(\boldsymbol{W},\boldsymbol{a})& :=\boldsymbol{G}_1(\boldsymbol{W})+\boldsymbol{G}_2(\boldsymbol{a}).\label{eq.G}
\end{align}
Let ${\cal W}$ denote the set of all  $\boldsymbol{W}$ satisfying (\ref{eq.p1b}), and ${\cal A}$ the set of all  $\boldsymbol{a}$ satisfying (\ref{eq.p1f}). We can then rewrite (\ref{eq.p1}) as
\begin{equation}\label{eq.compact}
\begin{split}
\min_{(\boldsymbol{W},\; \boldsymbol{a})}  ~~ &  f(\boldsymbol{W},\boldsymbol{a}) \\
\text{s. t.}  ~~ & \boldsymbol{G}(\boldsymbol{W},\boldsymbol{a}) \leq \boldsymbol{0}, \;\; \boldsymbol{W} \in {\cal W}, \;\; \boldsymbol{a} \in {\cal A}
\end{split}
\end{equation}

Note that functions $f(\boldsymbol{W},\boldsymbol{a})$ and $\boldsymbol{G}(\boldsymbol{W},\boldsymbol{a})$ in (\ref{eq.f})--(\ref{eq.G}) are linearly separable in $\boldsymbol{a}$ and $\boldsymbol{W}$, Hence, a Benders' decomposition approach can be developed to solve (\ref{eq.compact}). Relying on the concept of Benders' partitioning, we can deal with (\ref{eq.compact}) in $\boldsymbol{\alpha}$-space instead of $(\boldsymbol{W},\boldsymbol{a})$-space. In particular, we rewrite (\ref{eq.compact}) as:
\begin{equation}\label{eq.master}
    \min_{\boldsymbol{a}}\; v(\boldsymbol{a}), ~~~\text{s. t.}  ~~ \boldsymbol{a} \in {\cal A} \cap {\cal V}
\end{equation}
where
\begin{align}
    v(\boldsymbol{a})&:= \min_{\boldsymbol{W} \in {\cal W}} f(\boldsymbol{W},\boldsymbol{a}), \;\; \text{s. t.} \;\; \boldsymbol{G}(\boldsymbol{W},\boldsymbol{a}) \leq \boldsymbol{0}; \label{eq.v} \\
    {\cal V}&:=\{\boldsymbol{a}:\; \boldsymbol{G}(\boldsymbol{W},\boldsymbol{a}) \leq \boldsymbol{0} \text{ for some } \boldsymbol{W} \in {\cal W}\}.\label{eq.V}
\end{align}
Note that the set ${\cal V}$ contains all $\boldsymbol{a}$ for which the problem in (\ref{eq.v}) is feasible, and the set ${\cal A} \cap {\cal V}$ is in fact the $\boldsymbol{a}$-space projection of the feasible region with problem (\ref{eq.compact}).

The projected problem (\ref{eq.master}) is clearly equivalent to (\ref{eq.compact}). As both the function $v(\boldsymbol{a})$ and the set ${\cal V}$ are implicitly defined via (\ref{eq.v}) and (\ref{eq.V}), problem (9) is hard to be directly tackled. To solve it, we rely on a dual representation of ${\cal V}$ which is given by the intersection of a collection of regions containing this set. Introduce a dual variable vector $\boldsymbol{\lambda}:=[\lambda_1, \ldots, \lambda_L]  \in  \mathbb{R}^{L}$. According to \cite[Theorem 2.2]{Geo72}, we immediately have:
\begin{lemma}
A point $\boldsymbol{a} \in {\cal V}$ if and only if it satisfies the following (an infinite number of) constraints:
\begin{equation}\label{eq.lambda}
    \Bigl[\min_{\boldsymbol{W} \in {\cal W}} \boldsymbol{\lambda}^T\boldsymbol{G}(\boldsymbol{W},\boldsymbol{a})\Bigr] \leq 0, \quad \forall \boldsymbol{\lambda} \in \Lambda
\end{equation}
where the set $\Lambda:=\{\boldsymbol{\lambda}: \;\boldsymbol{\lambda} \geq \boldsymbol{0}, \text{ and } \boldsymbol{1}^T \boldsymbol{\lambda}=1\}$.
\end{lemma}

Lemma 1 in fact is a direct consequence of duality theory. Since the functions in $\boldsymbol{G}(\boldsymbol{W},\boldsymbol{a})$ are convex in $\boldsymbol{W}$, the problem (\ref{eq.v}) is convex for any given $\boldsymbol{a} \in {\cal A} \cap {\cal V}$. Let $\boldsymbol{\mu}:=[\mu_1, \ldots, \mu_L]  \in  \mathbb{R}^{L}$. By the strong duality between (\ref{eq.v}) and its dual problem, we can also mimic the proof of \cite[Theorem 2.3]{Geo72}\cite{Boyd04} to establish:
\begin{lemma}
For any $\boldsymbol{a} \in {\cal A} \cap {\cal V}$,
\begin{equation}
    v(\boldsymbol{a})=\max_{\boldsymbol{\mu}\geq \boldsymbol{0}} \Bigl[\min_{\boldsymbol{W} \in {\cal W}}\Big(f(\boldsymbol{W},\boldsymbol{a})+ \boldsymbol{\mu}^T\boldsymbol{G}(\boldsymbol{W},\boldsymbol{a})\Big)\Big].
\end{equation}
\end{lemma}

Based on Lemmas 1--2, we can then turn (\ref{eq.master}) into an equivalent form:
\begin{equation}
    \min_{\boldsymbol{a} \in {\cal A}}\Bigl\{\max_{\boldsymbol{\mu}\geq \boldsymbol{0}} \Bigl[\min_{\boldsymbol{W} \in {\cal W}}\Big(f(\boldsymbol{W},\boldsymbol{a})+ \boldsymbol{\mu}^T\boldsymbol{G}(\boldsymbol{W},\boldsymbol{a})\Big)\Big]\Big\}, \;\;\;\text{s. t.  \;\;(\ref{eq.lambda})}.\nonumber
\end{equation}
Introduce an auxiliary variable $a_0 \in \mathbb{R}$. The problem can be further reformulated as:
\begin{subequations}\label{eq.eqiv}
\begin{align}
& \min_{(\boldsymbol{a} \in {\cal A},\; a_0)} \; a_0 \label{eq.eqiva} \\
 \text{s.t.}~~& a_0 \geq \min_{\boldsymbol{W} \in {\cal W}}\Big(f(\boldsymbol{W},\boldsymbol{a})+ \boldsymbol{\mu}^T\boldsymbol{G}(\boldsymbol{W},\boldsymbol{a})\Big), \;\; \forall \boldsymbol{\mu} \geq \boldsymbol{0} \label{eq.eqivb}\\
  & \Bigl[\min_{\boldsymbol{W} \in {\cal W}} \boldsymbol{\lambda}^T\boldsymbol{G}(\boldsymbol{W},\boldsymbol{a})\Bigr] \leq 0, \quad \forall \boldsymbol{\lambda} \in \Lambda \label{eq.eqivc}
\end{align}
\end{subequations}

For convenience, we henceforth call (\ref{eq.eqiv}) a master problem. As there are infinitely many constraints in  the problem (14), a natural strategy to solve it is relaxation. Following Benders' decomposition approach, we can solve a relaxed version of (\ref{eq.eqiv}) ignoring all but a few constraints in the initial stage. If the returned solution cannot satisfy the ignored constraints, we select one of the violated constraints and add it to the relaxed problem, then solve the problem again. This continues until an optimal solution satisfying all the constraints is found, or a termination criterion is met.

A key step with the aforementioned approach is how to check the (in-)feasibility of a solution for a relaxed version of (\ref{eq.eqiv}) with respect to the ignored constraints and, if it is infeasible, how to select a violated constraint. Interestingly, this can be done by solving the problem in (\ref{eq.v}).

 From now on we refer to the problem in (\ref{eq.v}) for a given $\boldsymbol{a}$ as [(\ref{eq.v})--$\boldsymbol{a}$]. Given that $(\boldsymbol{\hat{a}}, \hat{a}_0)$ is an optimal solution for a relaxed version of (\ref{eq.eqiv}), it follows from the definition of ${\cal V}$ and Lemma 1 that $\boldsymbol{\hat{a}}$ satisfies (\ref{eq.eqivc}) if and only if the problem [(\ref{eq.v})--$\boldsymbol{\hat{a}}$] is feasible. In addition, if [(\ref{eq.v})--$\boldsymbol{\hat{a}}$] is feasible, then Lemma 2 infers that $(\boldsymbol{\hat{a}}, \hat{a}_0)$ satisfies (\ref{eq.eqivb}) if and only if $\hat{a}_0 \geq v(\boldsymbol{\hat{a}})$.

Indeed, the problem [(\ref{eq.v})--$\boldsymbol{\hat{a}}$] is a convex SOCP that admits efficient polynomial-time solver. Therefore, [(\ref{eq.v})--$\boldsymbol{\hat{a}}$] is suitable for checking the feasibility of $(\boldsymbol{\hat{a}}, \hat{a}_0)$, and any (primal-)dual type solver can produce an index of a violated constraint in case that $(\boldsymbol{\hat{a}}, \hat{a}_0)$ is infeasible. By an index of a violated constraint, we refer to a vector $\boldsymbol{\hat{\mu}}\geq \boldsymbol{0}$ such that
\begin{equation}\label{eq.inf1}
    \hat{a}_0 < \min_{\boldsymbol{W} \in {\cal W}}\Big(f(\boldsymbol{W},\boldsymbol{\hat{a}})+ \boldsymbol{\hat{\mu}}^T\boldsymbol{G}(\boldsymbol{W},\boldsymbol{\hat{a}})\Big)
\end{equation}
if (\ref{eq.eqivb}) is violated, or a vector $\boldsymbol{\hat{\lambda}} \in \Lambda$ such that
\begin{equation}\label{eq.inf2}
    \Bigl[\min_{\boldsymbol{W} \in {\cal W}} \boldsymbol{\hat{\lambda}}^T\boldsymbol{G}(\boldsymbol{W},\boldsymbol{\hat{a}})\Bigr] > 0
\end{equation}
if (\ref{eq.eqivc}) is violated. Actually, given that [(\ref{eq.v})--$\boldsymbol{\hat{a}}$] is infeasible, any dual-type solver would produce a non-zero $\boldsymbol{\tilde{\lambda}}$ satisfying (\ref{eq.inf2}). Then we are able to obtain the required $\boldsymbol{\hat{\lambda}} = \boldsymbol{\tilde{\lambda}}/\|\boldsymbol{\tilde{\lambda}}\|$, i.e., by normalizing $\boldsymbol{\tilde{\lambda}}$. Furthermore, given that [(\ref{eq.v})--$\boldsymbol{\hat{a}}$] is feasible and it has a finite optimal value, the dual-type solver can provide an $\boldsymbol{\tilde{\mu}}$ for its dual problem as a byproduct. By definition, we have
\begin{equation}
    \boldsymbol{\tilde{\mu}} = \arg \max_{\boldsymbol{\mu}\geq \boldsymbol{0}} \Bigl[\min_{\boldsymbol{W} \in {\cal W}}\Big(f(\boldsymbol{W},\boldsymbol{\hat{a}})+ \boldsymbol{\mu}^T\boldsymbol{G}(\boldsymbol{W},\boldsymbol{\hat{a}})\Big)\Big].
\end{equation}
If we have $\boldsymbol{\hat{\mu}}\geq \boldsymbol{0}$ satisfying (\ref{eq.inf1}), it must hold
\begin{align}
    \hat{a}_0 & < \min_{\boldsymbol{W} \in {\cal W}}\Big(f(\boldsymbol{W},\boldsymbol{\hat{a}})+ \boldsymbol{\hat{\mu}}^T\boldsymbol{G}(\boldsymbol{W},\boldsymbol{\hat{a}})\Big) \nonumber \\
    & \leq \min_{\boldsymbol{W} \in {\cal W}}\Big(f(\boldsymbol{W},\boldsymbol{\hat{a}})+ \boldsymbol{\tilde{\mu}}^T\boldsymbol{G}(\boldsymbol{W},\boldsymbol{\hat{a}})\Big).
\end{align}
This implies that $\boldsymbol{\tilde{\mu}}$ is an index of a violated constraint; indeed, it is the index for the most violated constraint. Therefore, we can set $\boldsymbol{\hat{\mu}} \equiv \boldsymbol{\tilde{\mu}}$.

In a nutshell, we show that [(\ref{eq.v})--$\boldsymbol{\hat{a}}$] can be used to check the feasibility of $(\boldsymbol{\hat{a}}, \hat{a}_0)$ for the master problem (\ref{eq.eqiv}), and to provide an index ($\boldsymbol{\hat{\lambda}}$ or $\boldsymbol{\hat{\mu}}$) of a (most) violated constraint in the case of infeasibility. This then enables development of the Benders' decomposition approach to solve (\ref{eq.eqiv}).

\subsection{Proposed Algorithm}

We next propose a Benders' decomposition algorithm to solve the intended joint BS activation and coordinated beamforming design problem. Define
\begin{align}
    & L^*(\boldsymbol{\alpha}, \boldsymbol{\mu}):=\min_{\boldsymbol{W} \in {\cal W}}\{f(\boldsymbol{W},\boldsymbol{a})+ \boldsymbol{\mu}^T\boldsymbol{G}(\boldsymbol{W},\boldsymbol{a})\} \nonumber \\
    & = \min_{\boldsymbol{W} \in {\cal W}}\{\sum_k \boldsymbol{w}_k^H \boldsymbol{w}_k + \boldsymbol{a}^T\boldsymbol{\pi} + \sum_l \mu_l ( \sum_k \boldsymbol{w}_k^H \boldsymbol{B}_l \boldsymbol{w}_k - a_l P_l) \}\nonumber \\
    & = \boldsymbol{a}^T (\boldsymbol{\pi}-\boldsymbol{\mu}\boldsymbol{P})+ C_1(\boldsymbol{\mu})
\end{align}
where $\boldsymbol{\mu}:=\{\mu_l,\forall l\}$, $\boldsymbol{P}:=\text{diag}(P_1, \ldots, P_L)$, and $C_1(\boldsymbol{\mu}):=\min_{\boldsymbol{W} \in {\cal W}}\{\sum_k \boldsymbol{w}_k^H \boldsymbol{w}_k+ \sum_l \mu_l\sum_k [\boldsymbol{w}_k^H \boldsymbol{B}_l \boldsymbol{w}_k]\}$.

If $\boldsymbol{\hat{\mu}}$ is an optimal dual vector for [(\ref{eq.v})--$\boldsymbol{\hat{a}}$], it follows that
\begin{align}
    L^*(\boldsymbol{a}, \boldsymbol{\hat{\mu}}) = \boldsymbol{a}^T (\boldsymbol{\pi}-\boldsymbol{\hat{\mu}}\boldsymbol{P})  + v(\boldsymbol{\hat{a}})-\boldsymbol{\hat{a}}^T (\boldsymbol{\pi}-\boldsymbol{\hat{\mu}}\boldsymbol{P}) ,
\end{align}
i.e., $C_1(\boldsymbol{\hat{\mu}})=v(\boldsymbol{\hat{a}})-\boldsymbol{\hat{a}}^T (\boldsymbol{\pi}-\boldsymbol{\hat{\mu}}\boldsymbol{P}) $, which can be really obtained after [(\ref{eq.v})--$\boldsymbol{\hat{a}}$] is solved.

In a similar way, let us define:
\begin{align}
    L_*(\boldsymbol{a}, \boldsymbol{\lambda})&:=\min_{\boldsymbol{W} \in {\cal W}} \boldsymbol{\lambda}^T\boldsymbol{G}(\boldsymbol{W},\boldsymbol{a}) \nonumber \\
    & = -\boldsymbol{a}^T \boldsymbol{\lambda}\boldsymbol{P} + C_2(\boldsymbol{\lambda})
\end{align}
where $C_2(\boldsymbol{\lambda}) := \min_{\boldsymbol{W} \in {\cal W}}\{\sum_l \lambda_l \sum_k [\boldsymbol{w}_k^H \boldsymbol{B}_l \boldsymbol{w}_k] \}$.

We now propose an efficient algorithm based on Benders' decomposition  to compute a solution $(\boldsymbol{W}^*, \boldsymbol{a}^*)$ for the problem (\ref{eq.prob})  of interest:
\begin{quote}
\vspace{0.05 in}\hrule height0.1pt depth0.3pt \vspace{0.05 in}
\begin{algorithm}
{\it Benders' decomposition method}

{\bf Initialize}: Given an accuracy level $\epsilon\geq0$ and a vector $\boldsymbol{\hat{a}} \in {\cal A}$, set $p=q=0$,
a lowerbound $\text{LB} = -\infty$, and a upperbound $\text{UB} = \infty$.

{\bf Repeat}: \\
    1) Solve [(\ref{eq.v})--$\boldsymbol{\hat{a}}$]; do according to one of the following two cases:
    \begin{itemize}
    \item[a)] Problem [(\ref{eq.v})--$\boldsymbol{\hat{a}}$] is infeasible: In this case, $\boldsymbol{\hat{a}} \notin {\cal V}$. Obtain a $\boldsymbol{\hat{\lambda}} \in \Lambda$ satisfying (\ref{eq.inf2}) and compute the function $L_*(\boldsymbol{a}, \boldsymbol{\hat{\lambda}})$. Let $q=q+1$, and $\boldsymbol{\lambda}_q = \boldsymbol{\hat{\lambda}}$; go to Step 2).

    \item[b)] Problem [(\ref{eq.v})--$\boldsymbol{\hat{a}}$] is feasible, i.e., $\boldsymbol{\hat{a}} \in {\cal V}$: Denote by $v(\boldsymbol{\hat{a}})$ and $\boldsymbol{W}(\boldsymbol{\hat{a}})$ the optimal value and optimal solution for [(\ref{eq.v})--$\boldsymbol{\hat{a}}$]. \\
        i) If $v(\boldsymbol{\hat{a}}) \leq \text{LB} + \epsilon$, terminate; output $\boldsymbol{W}^* =\boldsymbol{W}(\boldsymbol{\hat{a}})$, and $\boldsymbol{a}^*=\boldsymbol{\hat{a}}$.\\
        ii) Otherwise, determine the optimal multiplier $\boldsymbol{\hat{\mu}}$ and function $L^*(\boldsymbol{a}, \boldsymbol{\hat{\mu}})$. If $v(\boldsymbol{\hat{a}})< \text{UB}$, update $\text{UB}=v(\boldsymbol{\hat{a}})$, $\boldsymbol{W}^* =\boldsymbol{W}(\boldsymbol{\hat{a}})$, and $\boldsymbol{a}^*=\boldsymbol{\hat{a}}$. Let $\boldsymbol{\mu}_p = \boldsymbol{\hat{\mu}}$ and $p=p+1$; go to Step 2).
    \end{itemize}
    2) Solve the relaxed master problem:
    \begin{equation}\label{eq.relax}
    \begin{split}
        \min_{(\boldsymbol{a} \in {\cal A},\; a_0)}  ~~ &  a_0 \\
        \text{s. t.} ~~ ~~ & a_0 \geq L^*(\boldsymbol{a}, \boldsymbol{\mu}_j), \quad j=1,\ldots,p\\
        & L_*(\boldsymbol{a}, \boldsymbol{\lambda}_j) \leq 0, \quad ~j=1,\ldots,q
    \end{split}
    \end{equation}
 Denote by $(\boldsymbol{\hat{a}},\hat{a}_0)$ an optimal solution for (\ref{eq.relax}). If $\text{UB} \leq \hat{a}_0 + \epsilon$, terminate; output $(\boldsymbol{W}^*, \boldsymbol{a}^*)$. Otherwise, let $\text{LB}=\hat{a}_0$; return to Step 1).
\end{algorithm}
\vspace{0.05 in} \hrule height0.1pt depth0.3pt \vspace{0.1 in}
\end{quote}

In the proposed  Algorithm 1, we obtain $v(\boldsymbol{\hat{a}})$ in Step 1-b) by solving [(\ref{eq.v})--$\boldsymbol{\hat{a}}$] for a feasible $\boldsymbol{\hat{a}} \in {\cal A} \cap {\cal V}$. This provides an upperbound for the optimal value of the projected problem (\ref{eq.master}), which is in turn also an upperbound for that of the original problem (\ref{eq.prob}). Meanwhile, since $\hat{a}_0$ obtained in Step 2) is the optimal value of the equivalent master problem (\ref{eq.eqiv}) with some of its constraints removed, it certainly provides a lowerbound for the optimal value of (4). Clearly, the sequence of values for $\hat{a}_0$ obtained at successive executions of Step 2) is monotonically nondecreasing, as more and more constraints are added to the relaxed master problem (\ref{eq.relax}). Hence, the current $\hat{a}_0$ always gives the greatest lowerbound; that is why we can set $\text{LB}=\hat{a}_0$. However, the sequence of values for $v(\boldsymbol{\hat{a}})$ is not guaranteed to be monotonically non-increasing. Therefore, we need to compare and store the best known upperbound so far, i.e., the smallest $v(\boldsymbol{\hat{a}})$ found at all previous iterations into $\text{UB}$ in Step 1-b). Finally, when we have $v(\boldsymbol{\hat{a}}) \leq \text{LB} + \epsilon$ in Step 1-b) or $\text{UB} \leq \hat{a}_0 + \epsilon$ in Step 2), we actually have $\text{UB} \leq \text{LB}+\epsilon$. For these situations, a desired $\epsilon$-optimal solution $(\boldsymbol{W}^*, \boldsymbol{a}^*)$ is obtained for (\ref{eq.prob}).

In Step 2) of  Algorithm 1, we need an appropriate algorithm to solve (\ref{eq.relax}). Rewrite the problem more explicitly:
\begin{equation}
    \begin{split}
        \min_{(\boldsymbol{a},\; a_0)}  ~ &  a_0 \\
        \text{s. t.} ~ & \boldsymbol{a}^T (\boldsymbol{\pi}-\boldsymbol{\mu}_j\boldsymbol{P})
            \leq  a_0 - C_1(\boldsymbol{\mu}_j), \;\;\; j=1,\ldots,p\\
        & -\boldsymbol{a}^T \boldsymbol{\lambda}_j\boldsymbol{P} \leq  -C_2(\boldsymbol{\lambda}_j), \;\;\; j=1,\ldots,q\\
        & a_l \in \{0,1\}, \;\; \forall l
    \end{split}
\end{equation}
For a fixed $a_0$, the problem is in fact a binary integer feasibility problem which can be solved by e.g., Matlab bintprog function. By solving a series of such binary integer problems, we can utilize a bisection search to determine the optimal $\hat{a}_0$ and the corresponding optimal $\boldsymbol{\hat{a}}$.

In fact, the brand-and-bound method is also adopted by the binary integer program solvers (e.g., Matlab bintprog). However, different from the original mixed integer program (\ref{eq.p1}), the number of optimized variables in (\ref{eq.relax}) is greatly reduced; thus, the complexity  becomes affordable. In addition, the number of optimized variables for the subproblem [(\ref{eq.v})--$\boldsymbol{\hat{a}}$] becomes smaller as well; as a result, a reduced complexity is required in computing the optimal beamforming matrices $\boldsymbol{W}(\boldsymbol{\hat{a}})$ per iteration. This is exactly the motive power of Benders' partitioning method. Consequently, the proposed Algorithm 1 can obtain the joint BS activation and coordinated beamforming solution in an efficient manner.

\subsection{Finite Convergence}

To show the efficiency of the proposed Benders' decomposition approach, we formally establish that:
\begin{proposition}
For any $\epsilon \geq 0$, Algorithm 1 produces an $\epsilon$-optimal solution for (\ref{eq.prob}) in a finite number of iterations.
\end{proposition}


The proof is provided in Appendix \ref{A:Proof}, which mimics that of \cite[Theorem 2.4]{Geo72}. We include it for completeness. It is worth noting that Proposition 1 establishes the convergence of Algorithm 1 in a finite number of iterations even for $\epsilon =0$, i.e., when an exact optimal solution is pursued.

Proposition 1 states that the proposed algorithm may need to test all the points in the set ${\cal A}$  in the worst case. Such a complexity is clearly not affordable. Yet, our proposed algorithm continuously adds a most violated constraint to the relaxed master problem (\ref{eq.relax}), then optimally solves the problem to search the next candidate $\boldsymbol{\hat{a}}$. As a result, it can usually converge within a small number of iterations, given that the problem (\ref{eq.prob}) is feasible.

\section{A Low-Complexity Dual-subgradient Algorithm}

Based on the formulation (\ref{eq.p1}), a dual-subgradient based solver can be also developed. To this end, introduce the Lagrange multiplier vector $\boldsymbol{\lambda}:=[\lambda_1, \ldots, \lambda_L]$ associated with the constraints (\ref{eq.p1c}). The partial Lagrangian function is then
\begin{equation}
    L(\boldsymbol{W},\boldsymbol{a}, \boldsymbol{\lambda})=\sum_k \boldsymbol{w}_k^H \boldsymbol{w}_k + \boldsymbol{a}^T\boldsymbol{\pi} + \sum_l \lambda_l(\sum_k \boldsymbol{w}_k^H \boldsymbol{B}_l \boldsymbol{w}_k - a_l P_l)
\end{equation}
The Lagrange dual function is given by
\begin{equation}
    D(\boldsymbol{\lambda})=\min_{\boldsymbol{W} \in {\cal W}, \;\; \boldsymbol{a} \in {\cal A}}\; L(\boldsymbol{W},\boldsymbol{a}, \boldsymbol{\lambda})
\end{equation}
and the dual problem is
\begin{equation}\label{eq.dual}
    \max_{\boldsymbol{\lambda} \geq \boldsymbol{0}} \;D(\boldsymbol{\lambda}).
\end{equation}

To solve dual problem (\ref{eq.dual}), we can rely on the dual subgradient ascent based iteration
\begin{equation}\label{eq.subgradient}
    \lambda_l(j+1)=[\lambda_l(j) + s(j) g_{\lambda_l}(j)]^+, \quad \forall l
\end{equation}
where $j$ denotes the iteration index and $s(j)$ is an appropriate stepsize. The subgradient $\boldsymbol{g}_{\boldsymbol{\lambda}}(j):=[g_{\lambda_1}(j),\ldots, g_{\lambda_L}(j)]^T$ can be calculated by
\begin{equation}
    g_{\lambda_l}(j) = \sum_k \boldsymbol{w}_k^H(j) \boldsymbol{B}_l \boldsymbol{w}_k(j) - a_l(j) P_l
\end{equation}
where $\boldsymbol{W}(j):=\{\boldsymbol{w}_k(j)\}$ and $\boldsymbol{a}(j):=\{a_l(j)\}$ are given by
\begin{equation}\label{eq.sub1}
    \boldsymbol{W}(j) \in \arg \min_{\boldsymbol{W} \in {\cal W}} \; \sum_k \boldsymbol{w}_k^H (\boldsymbol{I} + \sum_l [\lambda_l(j)\boldsymbol{B}_l ])\boldsymbol{w}_k
\end{equation}
\begin{equation}\label{eq.sub2}
    \boldsymbol{a}(j) \in \arg \min_{\boldsymbol{a} \in {\cal A}} \sum_l ([\pi_l-\lambda_l(j) P_l] a_l)
\end{equation}

The subproblem in (\ref{eq.sub1}) is a standard SOCP; hence, $\boldsymbol{W}(j)$ can be computed by e.g., interior-point method in polynomial time \cite{Boyd04}. The subproblem in (\ref{eq.sub2}) is an integer linear program; an optimal $\boldsymbol{a}(j)$ can be found as:
\begin{equation}
    a_l(j) =
    \begin{cases}
    1, & \pi_l \leq \lambda_l(j) P_l\\
    0, & \pi_l > \lambda_l(j) P_l
    \end{cases}
\end{equation}

When we adopt a constant stepsize $s(j)=s$, the subgradient iterations (\ref{eq.subgradient}) can converge to a neighborhood (with its size proportional to stepsize $s$) of the optimal $\boldsymbol{\lambda}^*$ for the dual problem (\ref{eq.dual}) from any initial $\boldsymbol{\lambda}(0)$. Suppose that we adopt a sequence of non-summable and diminishing stepsizes satisfying $\lim_{j \rightarrow \infty} s(j) =0$ and $\sum_{j=0}^{\infty} s(j) = \infty$. Then the iterations (\ref{eq.subgradient}) can asymptotically converge to the exact $\boldsymbol{\lambda}^*$ as $j \rightarrow \infty$ \cite{Boyd04,Bertsekas2015}.

After the iterations (\ref{eq.subgradient}) converge to yield $\boldsymbol{\lambda}^*$, let
\begin{align}
    \boldsymbol{\overrightarrow{W}} \in \arg \min_{\boldsymbol{W} \in {\cal W}} \; \sum_k \boldsymbol{w}_k^H (\boldsymbol{I} + \sum_l [\lambda_l^*\boldsymbol{B}_l ])\boldsymbol{w}_k\\
    \boldsymbol{\overrightarrow{a}} \in \arg \min_{\boldsymbol{a} \in {\cal A}} \sum_l ([\pi_l-\lambda_l^* P_l] a_l)
\end{align}
Note that since the problem (\ref{eq.p1}) is nonconvex, there may exist nonzero duality gap; i.e., $(\boldsymbol{\overrightarrow{W}},\boldsymbol{\overrightarrow{a}})$ may not be a feasible solution for (\ref{eq.p1}). In this case, we simply use the BS activation vector $\boldsymbol{\overrightarrow{a}}$, and find the corresponding optimal beamforming matrix $\boldsymbol{\overrightarrow{W}}^*$ under such a BS activation situation; then output $(\boldsymbol{\overrightarrow{W}}^*,\boldsymbol{\overrightarrow{a}})$ as an approximate solution for (\ref{eq.p1}).
The proposed dual-subgradient algorithm is summarized as follows:

\begin{quote}
\vspace{0.05 in}\hrule height0.1pt depth0.3pt \vspace{0.05 in}
\begin{algorithm}
{\it Dual-subgradient approach}

{\bf Initialize}: select an initial $\boldsymbol{\lambda}(0)$, a stepsize $s$, an accuracy level $\epsilon\geq0$, and set $j=0$.

{\bf Repeat}: \\
    1) Solve (\ref{eq.sub1}) and (\ref{eq.sub2}), to obtain $\boldsymbol{W}(j),\boldsymbol{a}(j)$. \\
    2) Compute the subgradient $g_{\lambda}(j)$, then update the $\lambda(j+1)$ via (19). Check the condition $\frac{\|\lambda(j+1)-\lambda(j)\|}{\|\lambda(j)\|} \leq \epsilon$. If it is satisfied, let $\boldsymbol{\overrightarrow{W}}=\boldsymbol{W}(j) $ and $ \boldsymbol{\overrightarrow{a}}=\boldsymbol{a}(j)$, go to Step 3). Otherwise let $j=j+1$, go to Step 1).\\
    3) If $(\boldsymbol{\overrightarrow{W}}, \boldsymbol{\overrightarrow{a}})$ is feasible for (\ref{eq.p1}), output $(\boldsymbol{\overrightarrow{W}}, \boldsymbol{\overrightarrow{a}})$ as the solution. Otherwise, use the BS activation vector $\boldsymbol{\overrightarrow{a}}$ to find the corresponding optimal beamforming matrix $\boldsymbol{\overrightarrow{W}}^*$ and output $(\boldsymbol{\overrightarrow{W}}^*, \boldsymbol{\overrightarrow{a}})$  as the solution.
\end{algorithm}
\vspace{0.05 in} \hrule height0.1pt depth0.3pt \vspace{0.1 in}
\end{quote}

 In Algorithm 2, we need to solve a standard SOCP (20) with a worst-case complexity of $\cal{O}$$((LN_lK)^{3.5})$ and compute the solution (22) for subproblem (21) with a complexity of $\cal{O}$$(L)$ per iteration. It can be also shown that the proposed dual-subgradient iteration could converge geometrically fast under mild condition\cite{Bertsekas2015}.  Hence, the dual-subgradient based Algorithm 2 has a guaranteed polynomial-time computational complexity. This is different from the Benders' decomposition method in Algorithm 1, which has an exponential-time complexity in the worst case. Interestingly, simulation results in the sequel will show that such a low-complexity dual-subgradient method can yield a near-optimal solution with little performance loss.

\begin{remark}
Table I compares the complexity with the MIP, RMIP, as well as the proposed Benders' and dual-subgradient methods, where $L_{bd}$ and $L_{ds}$ represent the number of iterations required by Benders' and dual-subgradient method, respectively. For MIP, the built-in branch-and-bound approach requires $2^{L}$ times iterations in the worst case.
For RMIP, it needs $LK$ times iterations, leading to a high complexity when $K$ is large. Furthermore, the RMIP could incur a significant performance loss (e.g., approximate 5\% additional power consumption over the optimal benchmark, as shown in Fig. 4). The proposed Benders' decomposition algorithm can converge much faster than the MIP algorithm; i.e., $L_{bd}$ is usually a small number in many cases. The dual-subgradient based algorithm has the smallest complexity as it only needs to solve the problems (32) and (33) with a polynomial-time complexity $\cal{O}$$(L+(LN_lK)^{3.5})$ per iteration, and it could converge geometrically fast, i.e., in a small number $L_{ds}$ of iterations, to a near-optimal solution.
\end{remark}

\begin{table}[!htbp]
\centering
\caption{Computational Complexity Analysis}
\begin{tabular}{|c|c|c|}
\hline
Algorithms& The Complexity Order\\
\hline
MIP& $\cal{O}$($2^{L}(LN_lK)^{3.5}$)\\
\hline
RMIP& $\cal{O}$($(LK)(LN_lK)^{3.5}$)\\
\hline
Benders& $\cal{O}$($L_{bd}(2^{L}+(LN_lK)^{3.5})$)\\
\hline

dual-subgradient& $\cal{O}$($L_{ds}(L+(LN_lK)^{3.5})$)\\
\hline
\end{tabular}

\label{T:worstcase}
\end{table}
\section{Generalizations}

The proposed framework can be readily generalized to robust beamforming designs accounting for CSI errors, and to multi-cell HetNet setups performing partial coordinated transmissions.

\subsection{Robust Beamforming}

In practice, the CSI $\boldsymbol{h}_k$ is typically not possible to be precisely available at the central entity. With the help of past channel measurements and/or good channel predictions, an additive error model can be adopted: $\boldsymbol{h}_k = \tilde{\boldsymbol{h}}_k + {\boldsymbol{\delta}}_k$, where $\tilde{\boldsymbol{h}}_k$ is the channel estimation at the BS $k$. Suppose that channel uncertainty is bounded by a region \cite{Vucic2009, Luo2010, Song2012}:
\begin{equation}\label{channel uncertainities}
  \mathcal{H}_k:=\left\{\tilde{\boldsymbol{h}}_k + {\boldsymbol{\delta}}_k \ | \  \left\|{\boldsymbol{\delta}}_k \right\| \leq \xi_k \right\}, \quad \forall k,
\end{equation}
where $\xi_k>0$ specifies the radius of $\mathcal{H}_k$, and it is also assumed known.

Given the channel uncertainty region $\mathcal{H}_k$, we define the worst-case SINR with user $k$ as
\begin{equation}\label{eq.worst}
  \widetilde{\text{SINR}}_k := \min_{\boldsymbol{h}_k \in\mathcal{H}_k} \frac{|{\boldsymbol{h}_{k}}^H \boldsymbol{w}_k|^2}{\sum_{i\neq k} (|{\boldsymbol{h}_{k}}^H \boldsymbol{w}_i|^2) + \sigma_k^2}.
\end{equation}
To guarantee the quality-of-service, we require
\begin{equation}\label{eq.sinr}
    \widetilde{\text{SINR}}_k \geq \gamma_k, \quad \forall k
\end{equation}
With constraint (\ref{eq.probb}) replaced by (\ref{eq.sinr}), the problem (\ref{eq.prob}) is reformulated to pursue the robust beamforming design for coordinated downlink transmissions.

The problem (27) can be reformulated into a convex form using the well-known semidefinite program (SDP) relaxation technique. By the definitions of $\mathcal{H}_k$, the constraint $\widetilde{\text{SINR}}_k \geq \gamma_k$ can be rewritten as:
\begin{equation}\label{robustSINRconstraint}
F_k(\boldsymbol{\delta}_k)\geq 0~ \text{for all}~ \boldsymbol{\delta}_k ~\text{such that}~ {\boldsymbol{\delta}_k}^H\boldsymbol{\delta}_k \leq (\xi_k)^2,
\end{equation}
where
\begin{equation*}
   F_k(\boldsymbol{\delta}_k):= (\tilde{\boldsymbol{h}}_k+\boldsymbol{\delta}_k)^H(\frac{\boldsymbol{w}_k{\boldsymbol{w}_k}^H}{\gamma_k}-\sum_{i\neq k}\boldsymbol{w}_i{\boldsymbol{w}_i}^H)(\tilde{\boldsymbol{h}}_k+\boldsymbol{\delta}_k) -\sigma_k^2.
\end{equation*}

Define $\boldsymbol{X}_k:=\boldsymbol{w}_k {\boldsymbol{w}_k}^H$, which implicitly implies $\boldsymbol{X}_k\succeq 0$ and $\text{rank}(\boldsymbol{X}_k)=1$. By applying the celebrated S-procedure in robust optimization \cite{Polik2007} \cite[Appendix B.2]{Boyd04}, (\ref{robustSINRconstraint}) can be transformed into
 \begin{equation}\label{S-procedure-Cons}
 \begin{split}
 \boldsymbol{\Gamma}_k:=\left(\begin{matrix}
       \boldsymbol{Y}_k+\tau_k\boldsymbol{I} & \boldsymbol{Y}_k\tilde{\boldsymbol{h}}_k \\
       \tilde{\boldsymbol{h}}_k^{H}\boldsymbol{Y}_k^{H}   & \tilde{\boldsymbol{h}}_k^{H}\boldsymbol{Y}_k\tilde{\boldsymbol{h}}_k-\sigma_k^2-\tau_k\xi_k^2 \end{matrix} \right) \succeq 0,
\end{split}
\end{equation}
for a $\tau_k>0$ and
\begin{equation}
\boldsymbol{Y}_k:=\frac{1}{\gamma_k}\boldsymbol{X}_k-\sum_{i\neq k} \boldsymbol{X}_i.
\end{equation}

Let $\boldsymbol{X}=\{\boldsymbol{X}_k, \forall k\}$. Use $\boldsymbol{X}$ (instead of $\boldsymbol{W}$) as the optimization variables. Introduce auxiliary variables $\boldsymbol{\tau}:=\{\tau_k,\forall k\}$ and drop the rank constraints $\text{rank}(\boldsymbol{X}_k)=1$, $\forall k$. We can reformulate (\ref{eq.prob}) into:
\begin{subequations}\label{eq.probust}
\begin{align}
\min_{(\boldsymbol{X},\; \boldsymbol{\tau},\;\boldsymbol{a})} ~~& \sum_{k=1}^K \text{tr}(\boldsymbol{X}_k) + \boldsymbol{a}^T\boldsymbol{\pi} \label{eq.probusta} \\
 \text{s.t.}~~& \boldsymbol{\Gamma}_k \succeq 0, \;\; \boldsymbol{X}_k \succeq 0, \tau_k \geq 0, \quad \forall k \label{eq.probustb}\\
  & \sum_{k=1}^K \text{tr}(\boldsymbol{B}_l \boldsymbol{X}_k) \leq a_l P_l, \quad \forall l \label{eq.probustc} \\
  & a_l \in \{0,1\}, \;\; \forall l.
\end{align}
\end{subequations}

Let $\boldsymbol{\chi}:=\{\boldsymbol{X}, \boldsymbol{\tau}\}$. For a fixed $\boldsymbol{a}$, (\ref{eq.probust}) reduces to a convex SDP that can be solved in polynomial time. Also, we can write linearly separate functions $f(\boldsymbol{\chi},\boldsymbol{a})$ and $\boldsymbol{G}(\boldsymbol{\chi},\boldsymbol{a})$ in $\boldsymbol{\chi}$ and $\boldsymbol{a}$, then constitute a problem in a similar form to (\ref{eq.compact}). The proposed Benders' decomposition based and dual-subgradient based approaches can be then employed to solve this problem.

One issue is how to recover the optimal beamforming vectors $\boldsymbol{w}_k^*$ from $\boldsymbol{X}_k^*$ yielded by Algorithm 1 for (\ref{eq.probust}). If it happens that $\text{rank}({\boldsymbol{X}_k}^*)=1$, $\forall k$, then we clearly find the optimal beamforming vectors ${\boldsymbol{w}_k}^*$ for the original problem as the (scaled) eigenvector with respect to the only positive eigenvalue of ${\boldsymbol{X}_k}^*$. Given that the uncertainty bounds $\xi_k$  are sufficiently small, \cite[Theorem 1]{Song2012} established that S-procedure based SDP for downlink beamforming designs always has a rank-one optimal solution ${\boldsymbol{X}_k}^*$, $\forall k$. For large $\xi_k$ case, it cannot be proved that rank-one optimal solutions for (\ref{eq.probust}) always exist; in this case, we may adopt a randomized rounding strategy \cite{Luo2010} to obtain vectors ${\boldsymbol{w}_k}^*$ from ${\boldsymbol{X}_k}^*$ to nicely approximate the solution of the original problem. In fact, although no proof for rank-one solution in large $\xi_k$ case is available, it was observed in extensive simulations that this kind of SDPs can have a rank-one optimal solution in many cases \cite{Song2012}.

\subsection{Multi-Cell HetNet}

Consider a multi-cell HetNet with $M$ cells. For each cell $m=1, \ldots, M$, a set of ${\cal L}_m$ of BSs serve a set ${\cal K}_m$ of users. Each BS $l_m \in {\cal L}_m$ has $N_{l_m} \geq 1$ antennas whereas each user has a single antenna. The set of all BSs is then ${\cal L}=\bigcup_{m=1}^M {\cal L}_m$, and the set of all users is ${\cal K}=\bigcup_{m=1}^M {\cal K}_m$. In this setup, user $k_m \in {\cal K}_m$ can be only served by the BSs $l_m \in {\cal L}_m$, i.e., the BSs in its serving cell. In other words, only partial coordinated transmissions are allowed for serving users \cite{Hong13, Liao14}. Denote by $\boldsymbol{h}_{l_i,k_m} \in \mathbb{C}^{N_{l_i} \times 1}$ and $\boldsymbol{w}_{l_i,k_m} \in \mathbb{C}^{N_{l_i} \times 1}$ the channel vector and the transmit-beamforming vector from BS $l_i$ to user $k_m$, $\forall l \in {\cal L}_i$, $\forall k_m \in {\cal K}_m$, $\forall (i,k)$. 
Then the SINR for user $k_m \in {\cal K}_m$ is given by
\begin{equation}\label{eq.4b}
    \text{SINR}_{k_m} = \frac{\sum_{l_m \in {\cal L}_m} |\boldsymbol{h}_{l_m,k_m}^H \boldsymbol{w}_{l_m,k_m}|^2}{\sum_{k_i\neq k_m} \sum_{l_i \in {\cal L}_i}|\boldsymbol{h}_{l_i,k_m}^H \boldsymbol{w}_{l_i,k_i}|^2 + \sigma_{k_m}^2}.
\end{equation}
The transmit-power constraint with the BS $l_m \in {\cal L}_m$ is:
\begin{equation}\label{eq.4c}
    \sum_{k_m \in {\cal K}_m} \|\boldsymbol{w}_{l_m,k_m}\|^2 \leq a_{l_m} P_{l_m};
\end{equation}
and the total consumed power across all BSs is:
\begin{equation}\label{eq.4a}
    \sum_{m=1}^M \sum_{l_m \in {\cal L}_m} \sum_{k_m \in {\cal K}_m} \|\boldsymbol{w}_{l_m,k_m}\|^2 + \boldsymbol{a}^T \boldsymbol{\pi}.
\end{equation}

With the objective function in (\ref{eq.proba}), the constraints (\ref{eq.probb}), (\ref{eq.probc}) replaced by (\ref{eq.4a}), (\ref{eq.4b})--(\ref{eq.4c}), the resultant problem has the same structure as its single-cell version (\ref{eq.prob}). The proposed Benders' decomposition based and dual-subgradient based approaches then readily carry over to find the joint BS activation and beamforming design solution in this multi-cell setup.

With the SINR replaced by the worst-case SINR similarly as (\ref{eq.worst}), we can also formulate the multi-cell version of (\ref{eq.probust}). The proposed approaches can also carry over to yield the robust beamforming designs along with the optimal BS activation strategy for multi-cell HetNets.

\begin{figure}[t]
\centering \epsfig{file=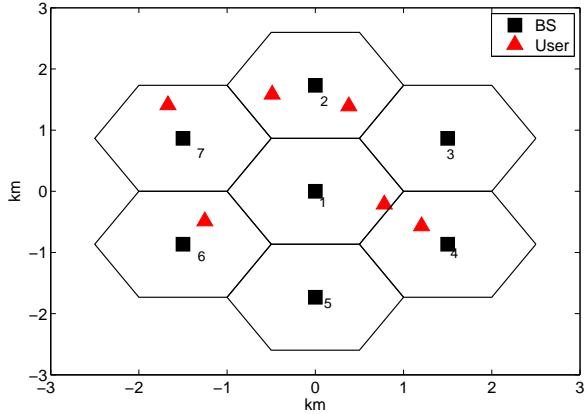, width=0.50\textwidth} 
\caption{A HetNet with $L=7$ BSs and $K=6$ users.  }
\label{F:System} 
\end{figure}

\section{Simulation Results}

Consider a cellular network comprising
7 identical hexagonal cells as shown in Fig. \ref{F:System}. We adopt similar channel and network power models as in \cite{Che13,Shi13}. Assume that one BS is located at each cell center. Each BS has two transmit antennas.   The   cell-radius is 1 kilometer (km), and the single-antenna users are uniformly distributed in the area.  The channel coefficient from the $m$-th antenna of the $l$-th BS to user $k$  is modeled as
\begin{equation}
h_{lk}^m =  \sqrt{L_{lk}(d_{lk}) \theta_{lk}  \xi_{lk}} \tilde{h}_{lk}^m,
\end{equation}
where $L_{lk}(d_{lk}) = 148.1 + 37.6\log_{10}(d_{lk})$ denotes the path loss at distance $d_{lk}$ (km), $\theta_{lk}=9$ dBi is the transmit antenna gain, $\xi_{lk}$ is the log-norm shadowing coefficient with zero mean and 8 dB variance, and $\tilde{h}_{lk}^m$ is small-scale Rayleigh fading coefficient. The noise variance is $\sigma_k^2 = -143$ dBm. The following power consumption model is adopted per BS:
\begin{equation}
P_{B,l} = \left\{
  \begin{array}{ll}
    P_{B,l}^{act} + \frac{1}{\eta} P_{B,l}^{tx}, & \hbox{{\text BS} {\it l} {\text is active};} \\
    P_{B,l}^{slp}, & \hbox{{\text BS} {\it l} {\text is not active}.}
  \end{array}
\right.
\end{equation}
where $P_{B,l}^{tx}$ denotes the transmit power, $P_{B,l}^{act}=6.8$ W stands for the implementation power consumption for the active BS, $\eta=25\%$ denotes the
  efficiency of the  power amplifier, and $P_{B,l}^{slp}=4.3$ W is the implementation power consumption for the BS
in the sleep mode. The maximum transmit power allowed by each BS is 43 dBm. Based on the above model, we have $\pi_l =  0.625$ for the optimization problem (3).

We compare the proposed algorithms with five baseline schemes, including random BS association (RBA), mixed integer programming (MIP), relaxed mixed integer programming (RMIP) \cite{Che13}, Sparsity based approach \cite{Liao14}, and joint BS association and power control (JBAPC) \cite{Qian13}\footnote{While the original JBAPC was developed for the  HetNet with single-antenna BSs, we generalize it to allow coordinated beamforming for the BSs equipped with multiple antennas.}.

\begin{figure}[t]
\centering \epsfig{file=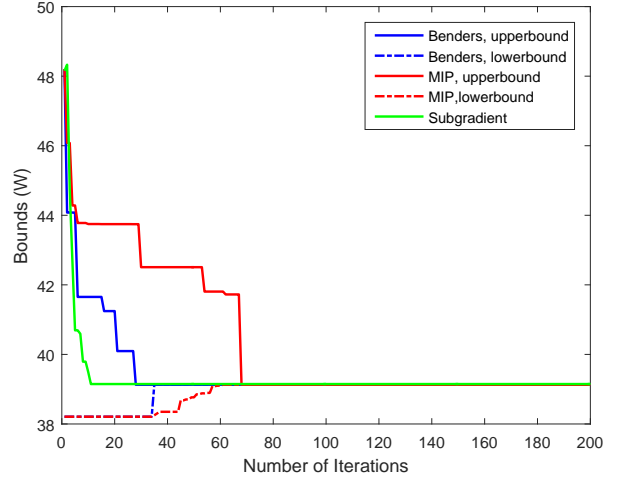,
width=0.50\textwidth} \vspace{-0.2cm} \caption{Convergence behaviors
of the proposed algorithms and the MIP for one channel realization when the SINR target is 15 dB. } \label{F:L7K6Convergence} \vspace{-0.2cm}
\end{figure}

For fair comparison, we first consider a HetNet with $L=7$ BSs and $K=6$ users.   Fig. \ref{F:L7K6Convergence} depicts the convergence behaviors of proposed Benders' decomposition based and dual-subgradient based algorithms as well as the MIP for one channel realization  when the SINR target is 15 dB and accuracy level $\epsilon = 0.01$. It is shown that
the proposed Benders' decomposition based scheme converges much faster than the MIP.  From simulation results, we find that 26 iterations are sufficient for the Benders' algorithm to find the optimal BS activation and coordinated beamforming, while more than 60 iterations are required for the MIP. In addition, the proposed dual-subgradient algorithm converges after only 12 iterations.
Note that the complexity with
the Benders' decomposition based algorithm can be smaller than that with the MIP per iteration since a SOCP (\ref{eq.v}) with a smaller size needs to be solved due to the Benders' partition. The dual-subgradient based algorithm has the smallest complexity as it only needs to solve the SOCP (20) in polynomial time, while the other two methods need to additionally solve an integer program, e.g., (17), with an exponential-time complexity in the worst case, in each iteration.

In a nutshell, the proposed Benders' decomposition based algorithm has significantly reduced computational complexity in finding the globally optimal solution than the MIP method. Furthermore, the proposed dual-subgradient algorithm has a very low (polynomial-time) complexity to yield a near-optimal solution with little performance loss. The latter performance loss is in fact due to the potential non-zero duality gap between the original (non-convex) problem (8) and its dual (29). Yet, it can be observed in Fig. 4 that the proposed dual-subgradient based scheme consumes only additional 0.5\% power on average  more than the optimal one for an SINR target of 5 dB.

\begin{figure}[t]
\centering \epsfig{file=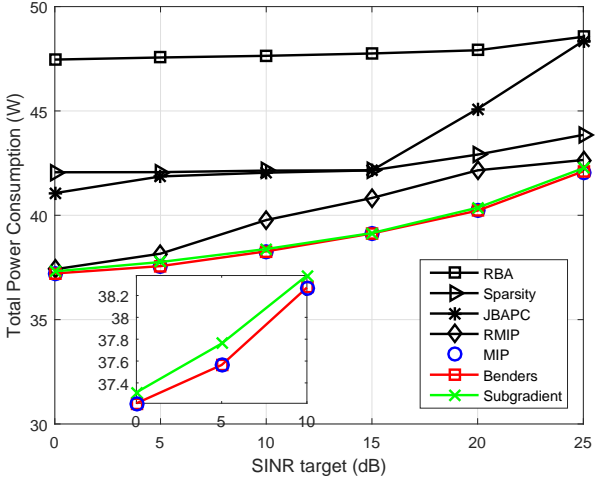,
width=0.50\textwidth} \vspace{-0.2cm} \caption{ Total power consumption
of a HetNet with $L=7$ BSs and $K=6$ users.}
\label{F:L7K6PowerNoCons} \vspace{-0.2cm}
\end{figure}

 Fig. \ref{F:L7K6PowerNoCons}  depicts the total power consumption with the RBA, sparsity based, JBAPC, RMIP, MIP, as well as the proposed Benders' and the dual-subgradient methods. Each point is computed by averaging over 100 channel realizations.  It is observed that
the total power consumption with the RBA, JBAPC, or sparsity-based algorithm  is
much higher than the optimal one provided by the MIP.
This is because coordinated beamforming is not considered in the JBAPC, whereas the sparsity based algorithm aims to also minimize the number of active BSs, leading to its sub-optimality.
The RMIP algorithm requires additional $10\%$ power consumption than the MIP.
It is also shown that the proposed Benders' decomposition based approach
always has the same total power consumption as the MIP for all
SINR targets. This corroborates the correctness of the Benders'
decomposition approach. With a much lower complexity, the proposed dual-subgradient based algorithm yields power consumption only slightly higher than the optimal one produced by the Benders' and MIP methods.

\begin{figure}[t]
\centering \epsfig{file=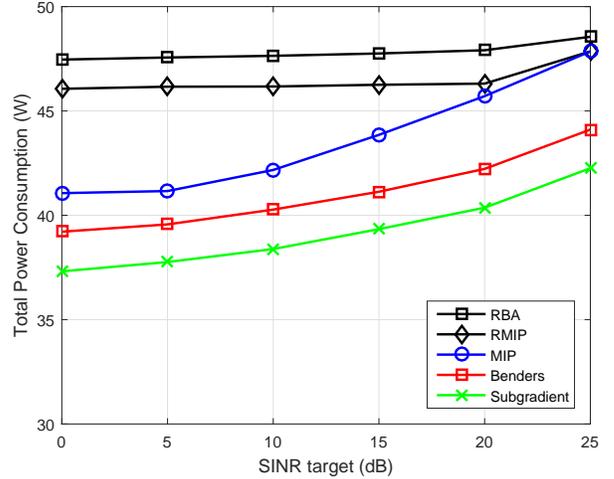, width=0.50\textwidth} 
\caption{ Total power consumption of a HetNet with $L=7$ BSs and $K=6$ users.  The maximum number of iterations for each algorithm is limited to be 16.}
\label{F:PowerL7K6Cons} 
\end{figure}

Next, consider the case with limitation on the maximum number of iterations to ensure an affordable computational complexity.  Fig. \ref{F:PowerL7K6Cons}  shows the total power consumption with the RBA, RMIP, MIP, Benders', and dual-subgradient methods where  the maximum number of iterations is limited to be 16. The other parameters are the same as in  Fig. \ref{F:L7K6PowerNoCons}.

 Note that due to large dimension of optimization variables, a large number of iterations could be required for the proposed Benders' decomposition approach and the MIP to converge. Yet, the proposed Benders' decomposition algorithm converges faster than the MIP algorithm.
 Compared to the results in Fig. \ref{F:L7K6PowerNoCons},
we see that when the number of iterations is limited, a much higher power budget is required for RMIP as well as MIP,
especially for high target SINR cases. It is also interesting to see   that the total consumed power with the MIP is much higher
 than that with the proposed Benders' decomposition algorithm. This is due to the fact that the standard branch-and-cut method  is quite inefficient and the MIP
 algorithm converges very slow in this medium-size network scenario.
More interestingly, the proposed dual-subgradient algorithm  always yields the minimum total power  among all the schemes,
capable of as large as $15\%$ power saving when compared to other algorithms for high-target SINRs. This indicates that when only low computational complexity is allowed, the proposed dual-subgradient algorithm can be an attractive candidate for finding an efficient BS activation and coordinated beamforming solution.

\begin{figure}[t]
\centering \epsfig{file=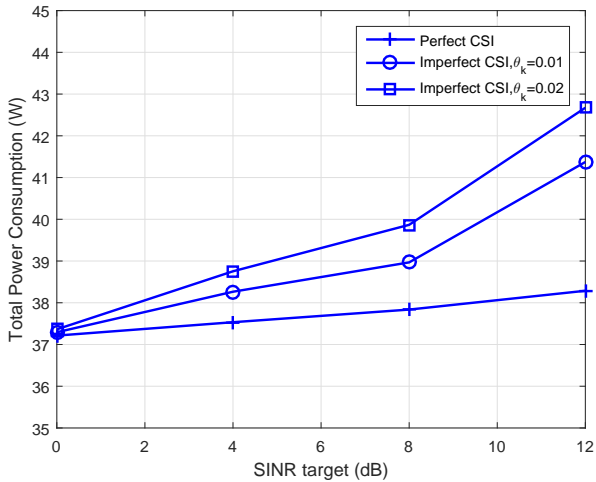, width=0.50\textwidth} 
\caption{ Total power consumption of a HetNet with uncertain CSI, with  $L=7$ BSs and $K=6$ users.}
\label{F:PowerImperfectCSI} 
\end{figure}

Fig. \ref{F:PowerImperfectCSI} depicts the total power consumption of a HetNet under channel uncertainty. There are $L=7$ BSs and $K=6$ users in the network. The channel uncertainty bounds are modeled as $\varepsilon_k = \theta_k \| \boldsymbol{h}_k \|, k=1,\dots,K$, where $\theta_k$ is chosen to be $\theta_k =0.01$ or 0.02 during simulations. The optimal BS activation and beamforming vectors are determined by the proposed Benders' decomposition based algorithm. It is observed that with imperfect CSI, more transmit power is required than that with perfect CSI. For instance, more than $10\%$ power is needed for an SINR target of 12 dB when $\theta_k = 0.02$.

\section{Conclusions}

We developed efficient optimal and suboptimal algorithms for joint BS activation and coordinated downlink
beamforming design in HetNets. While the proposed Benders' decomposition approach is capable of obtaining the global optimal solution
within a finite number of iterations, the proposed dual-subgradient scheme can yield a near-optimal solution with guaranteed very low (polynomial-time) complexity.  The simulated results validated that the proposed algorithms significantly outperform existing alternatives.

As green wireless communications have
received growing interest, some recent works have addressed the resource allocation for the smart-grid powered CoMP transmissions, where the BSs are jointly powered by persistent grid energy and harvested renewable energy sources \cite{XWang2015, XWang2016, XChen2017, XWang2018}. The energy harvesting (EH) communication integrated with smart grids clearly presents new theoretical and design challenges. Generalization of the proposed approaches to EH integrated smart-grid powered CoMP scenarios will be pursued in future research.

\appendices
  \section{Proof of the Proposition 1}
  \label{A:Proof}
    \begin{proof}
    For an arbitrary $\epsilon \geq 0$, finite termination directly follows from the finiteness of the set ${\cal A}$ as well as the fact that no $\boldsymbol{\hat{a}}$ can repeat itself in solution to (\ref{eq.relax}) in Step 2). This is because: i) if $\boldsymbol{\hat{a}} \notin {\cal V}$, then Step 1-a) generates a $L_*(\boldsymbol{a}, \boldsymbol{\hat{\lambda}})$ constraint that precludes $\boldsymbol{\hat{a}}$ from being feasible in (\ref{eq.relax}) ever again; ii) if $\boldsymbol{\hat{a}} \in {\cal V}$ and $v(\boldsymbol{\hat{a}})$ is finite, then a generated constraint $a_0 \geq L^*(\boldsymbol{a}, \boldsymbol{\hat{\mu}})$ would imply the $\epsilon$-optimality of $\boldsymbol{\hat{a}}$ if $\boldsymbol{\hat{a}}$ were ever to occur again as a solution of (\ref{eq.relax}) [if $(\bar{a}_0,\boldsymbol{\bar{a}}=\boldsymbol{\hat{a}})$ were to solve (\ref{eq.relax}) subsequently, then $\bar{a}_0 \geq L^*(\boldsymbol{\hat{a}}, \boldsymbol{\hat{\mu}}) = v(\boldsymbol{\hat{a}})=v(\boldsymbol{\bar{a}})$ would have to hold; hence, the termination condition would be satisfied].
    \end{proof}
%
%
%

\end{document}